# Compact fiber optical interferometer technique to measure picometer displacements in biological piezoelectric materials

Mingyue Liu, Nicholas Yaraghi, Jun Xu, David Kisailus and Umar Mohideen[a]

*Department of Physics and Astronomy, University of California, Riverside, California, 92521, USA*

**Abstract**

A simple and robust fiber optical interferometer was developed to non-invasively study the weak piezoelectric effect from thin samples. A biological sample from inter-molt dactyl clubs obtained from the mantis shrimp was used as the test sample. The non-contact technique can measure displacements better than 0.5 picometer for samples subjected to large electric fields. The approach utilizes the phase dependent detection of an oscillating cavity at different frequencies from 0.5 kHz to 2.0 kHz. The piezoelectric constant of the biological samples was calculated from the optical interference fringes and determined to be in the range of 0.3-0.5 pm/V. The noise of 20 fm/Hz$^{0.5}$ in the setup is primarily due to thermally associated strains from current flow to the sample electrodes during the measurement.

## 1. Introduction

Optical interferometric techniques provide unprecedented displacement resolution [1-11]. In general, optical interferometric approaches are non-invasive and contactless and thus ideal for measuring mechanical strain without need for special sample preparation related to the detection. They are particularly appropriate when a large electric field application is necessary for the effect to be studied such as in weak piezoelectric materials [12-14]. A piezoelectric material when subjected to an external electric field undergoes an extension or contraction depending on the direction of the applied field with respect to the alignment of the material polarization. The strain from the elongation or contraction of the piezoelectric material depends on the value of the electric field applied. By using non-contact optical techniques, large electric fields can be applied without affecting the strain measurement technique.

The fiber optic interferometric method developed here is compact and can reproducibly measure picometer displacements. As it is optical fiber based, it can be used with very small samples with surface dimensions smaller than a millimeter. With the displacement resolution of a picometer, very thin piezoelectric samples which undergo very small extensions and contractions on application of an electric field can be studied [15]. The technique will also find use in the calibration of micromachines and microactuators. As a test, we report on the measurement of the piezoelectric coefficient of a small and thin biological material.

Piezoelectricity in living systems is important both from a materials perspective as well as due to its role in biological growth. In fact the study of piezoelectricity in polymers began with the investigation of natural occurring materials [16]. Furthermore, it is known that the electric polarization in bone influences its growth [17,18]. For example, ambulatory motion in humans results in localized strain / deformation in bone, which is known to generate a small fluctuating current that stimulates local bone cell growth. In addition to bone, there are many other biological materials that have observed piezoelectric responses [19] including other collagen based materials such as tendon, teeth, skin and even the trachea. Keratin-based polymers such as hair and wool, as well as cellulose-based bio-materials including wood and bamboo show piezoelectric responses. Finally, chitin based bio-materials such as crab shell and lobster apodeme also demonstrate piezoelectric responses. As such, recent investigations [20-22] of an incredibly damage tolerant structure, the dactyl club (thoracic raptorial appendage used in smashing its prey) of a mantis shrimp (an aggressive marine crustacean) have suggested the potential for a piezoelectric response in the outer cuticle of its club. Given the biological significance this was chosen as the sample in this investigation.

---

[a] Electronic mail: umar.mohideen@ucr.edu

The paper is arranged as follows: In Section 2, we present a description of the optical interferometer based experimental setup for the phase sensitive detection of the displacement in the thin biological samples undergoing piezoelectric strain in response to an applied electric field. We also discuss the sample preparation and mounting. In Section 3, we discuss the method of measurement of the piezoelectric strain and the analysis used to determine the piezoelectric coefficient. In Section 4, readers will find the conclusion.

## 2. Experimental setup

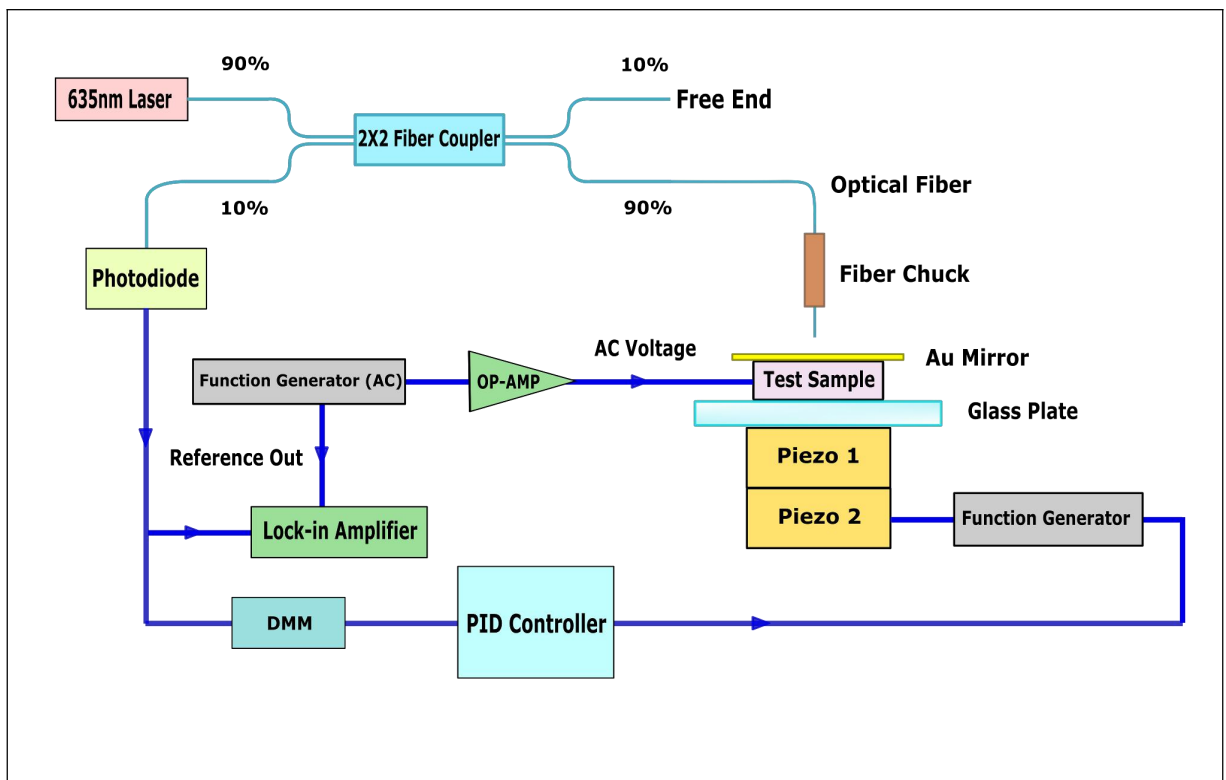

Figure 1. Schematic of the optical interferometer technique developed for the measurement of the very small piezoelectric strains. The specific components are labeled in the figure. Piezo 1 is used only for calibration purposes. Piezo 2 is used to maintain the cavity length at the Q-point.

Towards the goal of measuring picometer displacements necessary for investigating the piezoelectric coefficients of small and thin biological samples, a fiber optical based interferometer was developed. The measurement principle is based on optical interferometry. A schematic of the experimental setup is shown in figure 1. A pigtailed Fabry-Perot Benchtop Laser Source (S1FC635, Thorlabs, Inc., USA) with an emission wavelength of 635 nm and output power of 0.05 mW was used as the light source. The output of the laser source was coupled to a single mode 2 × 2 fused fiber optic coupler with a split ratio 90:10 (Thorlabs, Inc., USA). The 90 percent end of the output fiber was cleaved to expose a clean perpendicular face. This forms one end of the interferometer cavity as shown in figure 2. The second surface of the Fabry-Perot interferometer is an Au coated silicon plate. The Au coating was made on the polished side of a 1 cm × 1 cm silicon wafer of 500 µm thickness. The cleaved end of the optical fiber and the Au mirror formed the two surfaces of a low finesse optical cavity. The biological sample to be investigated was attached to the bottom of the silicon plate using epoxy and its preparation is described later. The output light of the interferometer was collected and transmitted through the same fiber and was detected by a photodiode (2107-FC-M, Newport Co.,USA) as shown in figure 1. This output light of the interferometer at the photodiode can be expressed as:

$$I_{out} = (\vec{E}_1 + \vec{E}_2)^* \cdot (\vec{E}_1 + \vec{E}_2), \qquad (1)$$

Where $\vec{E}_1$ and $\vec{E}_2$ represent electric fields of reflected beams from the cleaved fiber end and the Au mirror respectively. This can be rewritten in terms of the corresponding light intensities involved and the cavity length $d$ as:

$$I_{out} = I_{effective}[1 + V\cos(\frac{4\pi n d}{\lambda})], \qquad (2)$$

where $I_{effective} = I_{input}(R_1 + \alpha(d)R_2)$ and the visibility $V = 2\sqrt{R_1 R_2 \alpha(d)}/(R_1 + \alpha(d)R_2)$. Here $I_{input}$ is the input intensity into the cleaved end of the fiber, $R_1$, $R_2$ are power reflection coefficients of the cleaved fiber end and the Au mirror, $\alpha(d)$ is the attenuation coefficient of the optical intensity due to the divergence of the light beam from the fiber, $n$ is index of refraction of air, and $\lambda$ = 635 nm is the wavelength of the red laser. $I_{effective}$ and visibility $V$ can be considered independent of $d$ when the changes in the interferometer cavity length are small.

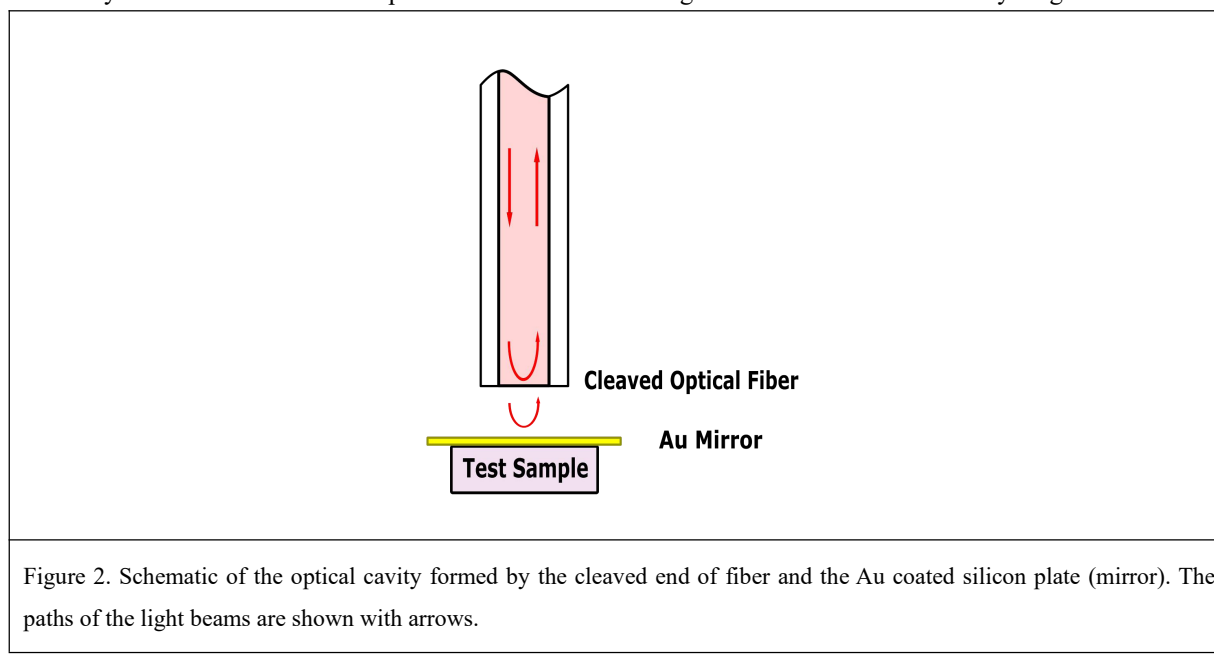

Figure 2. Schematic of the optical cavity formed by the cleaved end of fiber and the Au coated silicon plate (mirror). The paths of the light beams are shown with arrows.

Our test samples consist of fresh inter-molt dactyl clubs obtained from both live and recently deceased mantis shrimp (Odontodactylus scyllarus, crustacean). Polished cross-sections of the dactyl club exocuticle were prepared by embedding in epoxy resin (System 2000, Fibreglast, USA), sectioning along the exocuticle-endocuticle interface in the coronal plane using a low-speed saw with diamond blade (TechCut 4, Allied High Tech Products, Inc., USA), and polishing on either side with progressively finer silicon carbide and diamond abrasive down to 50 nm grit. Samples were washed in deionized (DI) water between polishing steps and subsequently sonicated in DI water for 15 to 30 minutes to remove any embedded polishing media. Epoxy resin surrounding the samples was carefully removed using a razor blade and the polished sections were dried in air prior to use. A polished biological sample of size 5 mm × 3 mm and thickness 0.5 mm was used. To facilitate the application of voltages to the biological sample, the top and bottom surfaces were made conductive with Ag epoxy (MG chemicals, USA). This prepared sample was then attached to the bottom of the Au coated silicon mirror as shown in figure 1.

To allow for control and modulation of the interferometer cavity length, a special arrangement of two Lead Zirconate Titanate (PZT) piezo blocks (Thorlabs, Inc., USA) of size 0.3 cm × 0.3 cm × 0.2 cm were used. In order to electrically isolate the sample from the piezo blocks beneath, a glass plate with 1 mm thickness was inserted between them. The two piezo blocks were mounted below the glass plate as shown in figure 1. A non-conductive cyanoacrylate epoxy was used to attach the glass plate as well as the piezo blocks. The whole experimental setup was built on an optical table to provide the necessary isolation from mechanical vibrations.

### 3. Measurement procedure & results

As a demonstration of the instrument development, a report of the piezoelectric coefficient $d_{33}$ of a biological sample is presented below. The test samples used were prepared as described above from fresh inter-molt dactyl clubs obtained from mantis shrimps. The biological sample piezoelectric coefficient was determined by comparing to the independently calibrated PZT piezo 1. First the interferometer signal was optimized by bringing the optical fiber within 20 μm of the Au mirror using a linear translation stage. The calibration of piezo 1 was completed as follows. A 0.1 Hz triangular wave with an amplitude of 10 V was applied to piezo 1. This applied voltage value was sufficient to generate more than 300 nm displacement of the Au mirror. The interferogram shown in figure 3 was generated using the triangular wave. The peaks correspond to constructive interference maxima and the valleys to destructive interference minima. The interference output is sinusoidal due to the low finesse of the cavity which in turn results from the use of the cleaved end of the optical fiber as one surface. The interferogram is fit to equation (2) using $d=\alpha V$ where $\alpha$ is the expansion coefficient and $V$ is the voltage applied to the piezo 1. The mean value found for the $\alpha$ of piezo 1 from the fit was 19.2 nm/V.

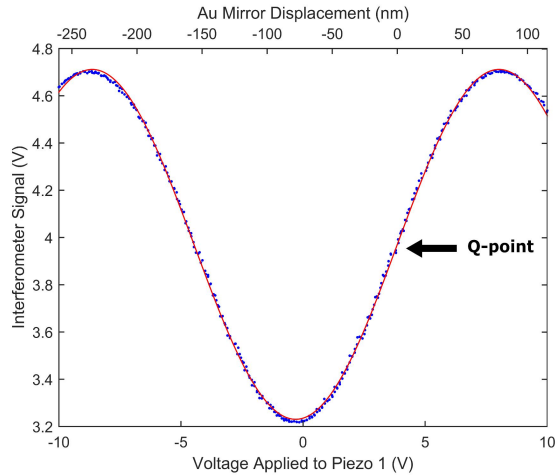

Figure 3. The interferometer output signal is a sinusoidal function of the distance moved by the Au mirror using piezo 1. The red solid line is the best fit with equation (2) using $d = \alpha V$ to determine the expansion coefficient $\alpha$ of piezo 1. This $\alpha$ is used to get the displacement of piezo 1 shown in the top axis. The signal has linear response at the Q-point as shown in the figure.

To maximize the sensitivity of the interferometer to very small changes in the cavity length, the Au mirror needs to be positioned such that the cavity length corresponds to the midpoint between an interference maxima and minima, hereafter referred to as the Q-point. In the experimental setup in figure 1, piezo 2 is used to maintain the cavity length at the Q-point using applied voltages. To find the Q-point, a 0.1 Hz triangular voltage waveform with an amplitude of 10 V was applied to piezo 2. An interference pattern similar to figure 3 was generated. The piezo 2 position and applied voltage corresponding to the Q-point can be identified. Once the Q-point was located, the corresponding DC voltage was applied to piezo 2 to bring the optical cavity to the appropriate length. A PID (proportional–integral–derivative) feedback system with a response frequency of 1 Hz was used on the piezo 2 voltage to keep the cavity length at the Q-point. The use of the PID feedback compensated any effects of mechanical drift in the components that form the optical cavity. Around the Q-point, the output intensity $I_{out}$ has a linear response to small change in the cavity length $d$.

The sensitivity was improved with phase sensitive detection using a lock-in amplifier. When sinusoidaly oscillating voltages were applied to the biological sample, the associated oscillating piezo extension in turn leads to an oscillating optical cavity length $d$. The corresponding oscillation in the interference signal was detected by the photodiode and its output was measured using a lock-in amplifier. As the fiber-plate separation was maintained at the Q-point, the oscillating optical signal generated was linearly proportional to the mechanical vibration of the biological piezoelectric sample. The same signal generator applying the sinusoidal voltage signal to the biological sample was used as the reference for the lock-in amplifier. For the particular measurements of the piezoelectric coefficient $d_{33}$ of the biological sample reported here, a sinusoidal voltage at 500 Hz and an amplitude of 35 V was first applied using a function generator. The amplitude of the function generator was amplified using an operational amplifier. Considering the extremely low piezoelectric coefficient of the biological sample, large voltages were necessary to get measurable signals. The output amplitude of the lock-in amplifier was recorded. The experiment was repeated 5 times and the average recorded lock-in amplitude is shown in figure 4 as a solid square. The error bar is too small to be displayed in the figure. The measurement was repeated for different voltage amplitudes of 70 V, 105 V, 130 V, 140 V and the average recorded values are shown in figure 4. To eliminate ambiguities from stray electrostatic fields interacting with the fiber and leading to its displacement, the voltage connections to the two electrodes were interchanged and the experiment repeated. There was no change in the response of the biological sample to this interchange.

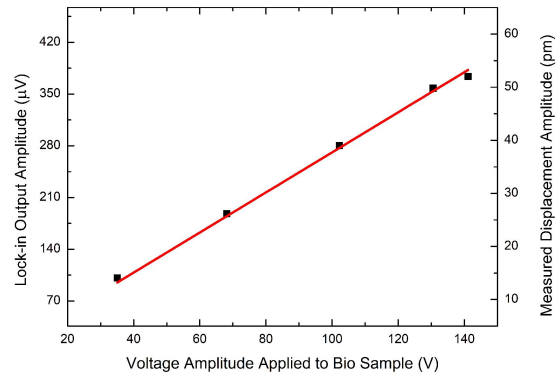

Figure 4. The lock-in amplitudes shown for the different voltage amplitudes applied to the biological piezoelectric sample. The data points are average of 5 measurements. The lock-in amplitude is converted to the corresponding displacement shown on the right axis using the calibrated piezo 1 data from figure 5 below. The slope of the best fit line leads to the piezoelectric coefficient $d_{33}$ = 0.377 pm/V at 500 Hz.

The output of the lock-in amplifier was calibrated using the earlier determined expansion coefficient α of piezo 1 using the interferometer. Here, the biological sample surfaces were electrically grounded. The cavity length was maintained at the same Q-point as before. Next, sinusoidal voltages with the frequency of 500 Hz and an amplitude of 4 mV was applied to piezo 1 using the same function generator. The amplitude output signal of the lock-in amplifier was recorded. Again, 5 measurements were taken and the average value is shown in figure 5 as a solid square. The measurements were repeated with 500 Hz signals of 6.0 mV, 8.0 mV, 10.0 mV applied to piezo 1. The recorded lock-in average amplitude values are shown in figure 5 for these applied voltage amplitudes. The error bars are also too small to be displayed in the figure. Using the interferometer calibrated piezoelectric coefficient of 19.2 nm/V determined earlier, the voltage value in the bottom axis can be converted to the corresponding expansion of the piezo 1 which is shown along the top axis in figure 5. A best fit straight line is fit to the data and the lock-in output signal per unit expansion of piezo 1 is determined to be 7.19 ± 0.03 mV/nm.

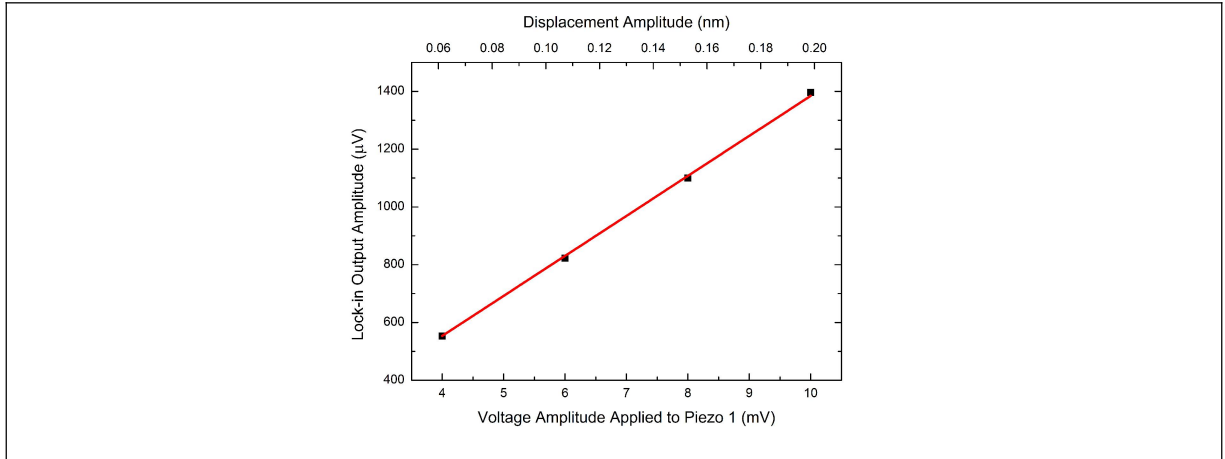

Figure 5. Calibration of the lock-in amplitude output using the calibrated expansion of piezo 1. The measured lock-in amplitudes for the various sinusoidal voltage amplitudes applied to piezo 1 are shown as solid squares. The top axis is the corresponding expansion amplitude obtained from the calibrated $\alpha$ of piezo 1 in figure 3. This measurement was taken at a frequency of 500 Hz.

The above calibrated value of the lock-in output amplitude can be used with that found in figure 4 to get the piezoelectric strain oscillation of the biological sample to the applied voltages. The value of the lock-in output in the left axis of figure 4 is converted to an expansion in nanometer using the above calibration and is shown on the right axis of figure 4. The slope of the best fit line is the $d_{33}$ piezoelectric coefficient of the biological sample and is found to be 377 ± 4 fm/V. The experiments were repeated for different sinusoidal frequencies of 1.0 kHz, 1.5 kHz and 2.0 kHz. The corresponding piezoelectric coefficients obtained for the biological sample are shown in table 1. The experiment was repeated using a second biological sample prepared using the same procedure.

Table 1: Piezoelectric coefficients of the biological samples. Column 1 has the frequencies of the applied voltages; column 2 and 3 are piezoelectric coefficients of the two biological samples measured. Column 4 reports the values for glass which is a non-piezoelectric material and is thus a measure of the background noise in the setup.

| Frequency (kHz) | Biological Sample 1 (pm/V) | Biological Sample 2 (pm/V) | Glass (pm/V) |
| --- | --- | --- | --- |
| 0.5 | 0.337 | 0.468 | 0.039 |
| 1.0 | 0.401 | 0.329 | 0.025 |
| 1.5 | 0.439 | 0.299 | 0.045 |
| 2.0 | 0.408 | 0.356 | 0.024 |

To estimate the background noise additional tests were performed. The biological piezoelectric sample was replaced by a glass sample of similar thickness and the noise background of the setup was measured. For this, a 1 × 1 cm piece of polished silicate glass was used. Since glass is centrosymmetric, it has zero piezoelectric response. It does have a second order size change under an applied electric field due to electrostriction, which is extremely small and of the order of $10^{-20}$ nm/V$^2$ [23]. Thus, any measured displacement signal on application of the voltage to the glass plate is the background noise. Figure 6 is a demonstration of the background noise signal measured as a function of the amplitude of the applied voltage. This observed background noise signal is probably due to thermally induced mechanical strains in the electrodes from the current flow during to the application of voltages. From the error bars in figure. 6, the displacement sensitivity of the interferometer is shown to be 0.5 picometer for the frequencies used. The bandwidth used in all measurements is 530 Hz from the the time constant setting of the lock-in amplifier used. Based on the noise level of the interferometer setup is 20 fm/Hz$^{0.5}$.

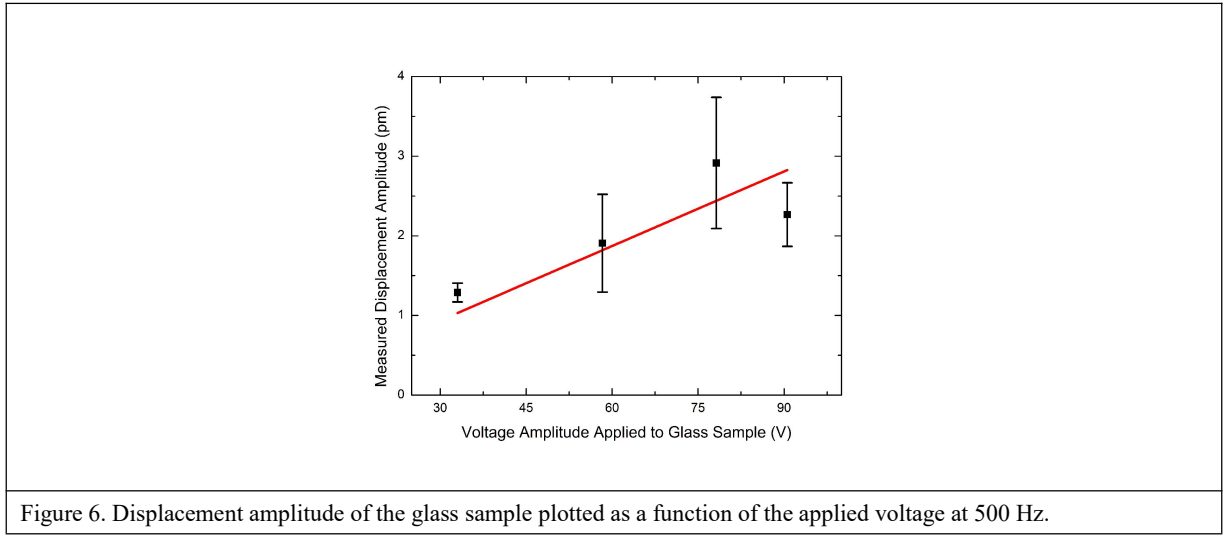

Figure 6. Displacement amplitude of the glass sample plotted as a function of the applied voltage at 500 Hz.

## 4. Conclusion

An optical interferometric system using phase sensitive detection was developed that can enable the non-contact measurement of very small piezoelectric strains in small samples. The experimental platform is versatile and can be used for measuring the piezoelectric coefficients of very thin samples with weak piezoelectric responses under large electric fields. We used the experimental setup to measure the piezoelectric coefficients of small and thin biological samples from the dactyl clubs of mantis shrimps. The noise and displacement sensitivity of the interferometer is shown to be 20 fm/Hz$^{0.5}$ and 0.5 picometer respectively. A non-piezoelectric material such as glass was used to measure the background noise in the system. The technique developed is simple and robust and will find other uses in non-invasive measurements of micro actuators and miniature or thin film piezoelectric samples as it exhibits high sensitivity and displacement resolution over a wide range of frequency.

**Acknowledgments**

Mingyue Liu, Jun Xu and Umar Mohideen were partially supported by NSF Grant No. PHY-1607749. The authors thank Robert Schafer and Shaolong Chen for discussions.